\documentclass[usenatbib]{mn2e}
\usepackage{graphicx}
\usepackage{amssymb}
\usepackage{color}

\newcommand{\be}{\begin{equation}}
\newcommand{\ee}{\end{equation}}
\newcommand{\ba}{\begin{eqnarray}}
\newcommand{\ea}{\end{eqnarray}}
\newcommand{\lp}{\left(}
\newcommand{\rp}{\right)}
\newcommand{\lb}{\left[}
\newcommand{\rb}{\right]}

\newcommand{\densnuc}{\rho_{\mathrm{nuc}}}
\newcommand{\Tc}{T_{\mathrm{c}}}
\newcommand{\Tcj}{T_{\mathrm{cx}}}
\newcommand{\Lr}{L_r}
\newcommand{\Ts}{T_{\mathrm{s}}}
\newcommand{\Qheat}{Q_{\mathrm{h}}}
\newcommand{\Qheati}{Q_{\mathrm{0}}}
\newcommand{\Lheat}{L_{\mathrm{h}}}
\newcommand{\densheat}{\rho_{\mathrm{h}}}
\newcommand{\Deltaheat}{\Delta_{\mathrm{h}}}
\newcommand{\tauheat}{\tau_{\mathrm{h}}}
\newcommand{\heatcap}{C}
\newcommand{\heatcapj}{C_{\mathrm{x}}}

\newcommand{\heatcapsfj}{\heatcap_{\mathrm{x}}^{\mathrm{SF}}}

\newcommand{\redheatcapj}{R^\heatcap_{\mathrm{x}}}
\newcommand{\kb}{k}
\newcommand{\kdiff}{K}
\newcommand{\pf}{p_{\mathrm{Fx}}}
\newcommand{\kf}{k_{\mathrm{Fx}}}
\newcommand{\kfn}{k_{\mathrm{F}n}}
\newcommand{\kfp}{k_{\mathrm{F}p}}
\newcommand{\meff}{m_{\mathrm{x}}^\ast}
\newcommand{\meffe}{m_{\mathrm{e}}^\ast}
\newcommand{\Gammac}{\Gamma_{\mathrm{c}}}
\newcommand{\Tb}{T_{\mathrm{env}}}
\newcommand{\densb}{\rho_{\mathrm{env}}}
\newcommand{\sgrav}{g_{\mathrm{s}}}
\newcommand{\sgravonefour}{g_{\mathrm{s,14}}}
\newcommand{\Tcns}{T_{\mathrm{cns}}}
\newcommand{\Tcnt}{T_{\mathrm{cnt}}}
\newcommand{\Tcp}{T_{\mathrm{cp}}}
\newcommand{\Tmelt}{T_{\mathrm{melt}}}
\newcommand{\Tdebye}{T_{\mathrm{D}}}
\newcommand{\Tmag}{T_{\mathrm{B}}}
\newcommand{\densmag}{\rho_{\mathrm{B}}}
\newcommand{\tdecay}{\tau_{\mathrm{decay}}}
\newcommand{\tamb}{\tau^{\mathrm{amb}}}
\newcommand{\tambir}{\tau^{\mathrm{nsol}}}
\newcommand{\tambssc}{\tau^{\mathrm{sol}}_{\mathrm{sc}}}
\newcommand{\tambirsc}{\tau^{\mathrm{nsol}}_{\mathrm{sc}}}
\newcommand{\Ttr}{T^{\mathrm{tr}}}
\newcommand{\Ttrsc}{\Ttr_{\mathrm{sc}}}
\newcommand{\ttamb}{\tau^{\mathrm{amb}}_{\mathrm{sc}}}

\title[Superfluid core in hot magnetars]{Magnetars: Super(ficially) hot and
super(fluid) cool}
\author[Ho, Glampedakis, \& Andersson]{Wynn C. G. Ho,$^1$\thanks{Email:
 wynnho@slac.stanford.edu}
Kostas Glampedakis,$^{2,3}$ and Nils Andersson$^1$
\\
$^1$School of Mathematics, University of Southampton, Southampton, SO17 1BJ \\
$^2$Departamento de F\'{i}sica, Universidad de Murcia, E-30100 Murcia, Spain \\
$^3$Theoretical Astrophysics, University of T\"{u}bingen,
Auf der Morgenstelle 10, T\"{u}bingen D-72076, Germany
}
\date{Accepted 2012 February 25. Received 2012 February 24;
 in original form 2011 November 30}

\begin{document}
\pagerange{\pageref{firstpage}--\pageref{lastpage}} \pubyear{2012}

\maketitle

\label{firstpage}

%%%%%%%%%%%%%%%%%%%%%%%%%%%%%%%%%%%%%%%%%%%%%%%%%%%%%%%%%
\begin{abstract}
We examine to what extent the inferred surface temperature of magnetars in
quiescence can constrain the presence of a superfluid in the neutron star
core and the role of magnetic field decay in the core.
By performing detailed simulations of neutron star cooling,
we show that extremely strong heating from field decay in the {\it core}
cannot produce the high observed surface temperatures
nor delay the onset of neutron superfluidity in the core.
We verify the results of Kaminker et al., namely that the high magnetar
surface temperatures require heating in the neutron star {\it crust},
and crust heating is decoupled from cooling/heating in the core.
Therefore, because crust heating masks core heating, it is not possible
to conclude that magnetar cores are in a non-superfluid state purely
from high surface temperatures.
From our interior temperature evolutions and after accounting for proton
superconductivity in the core, we find that neutron superfluidity in the
core occurs less than a few hundred years after neutron star formation.
This onset time is unaffected by heating due to core field
decay at fields $\lesssim 10^{16}\mbox{ G}$.
Thus all known neutron stars, including magnetars, without a core containing
exotic particles, should have a core of superfluid neutrons and
superconducting protons.
\end{abstract}


\maketitle

%%%%%%%%%%%%%%%%%%%%%%%%%%%%%%%%%%%%%%%%%%%%%%%%%%%%%%%%%%%%%%%%%%%
\section{Introduction} \label{sec:intro}

Neutron stars (NSs) begin their lives very hot (with temperatures
$T>10^{11}\mbox{ K}$) but cool rapidly through the emission of neutrinos.
Neutrino emission processes depend on
uncertain physics at the supra-nuclear densities
($\rho>\densnuc\approx 2.8\times 10^{14}\mbox{ g cm$^{-3}$}$) of the NS core
(see \citealt{tsuruta98,yakovlevpethick04,pageetal06}, for review).
Current theories indicate that the core may contain a neutron superfluid
and proton superconductor or even exotic particles,
such as hyperons and deconfined quarks
(see, e.g., \citealt{lattimerprakash04,haenseletal07}, for review).
The recent observation of rapid cooling \citep{heinkeho10,shterninetal11}
of the NS in the Cassiopeia~A supernova remnant
provides the first constraints on the critical temperatures for the onset
of superfluidity of core neutrons $\Tcnt$ (in the triplet state) and protons
$\Tcp$ (in the singlet state),
i.e., $\Tcnt\approx(5-9)\times 10^8$~K and $\Tcp\sim (2-3)\times 10^9$~K
\citep{pageetal11,shterninetal11}.

Anomalous X-ray pulsars and soft gamma-ray repeaters form the magnetar
class of NSs, i.e., systems which possess superstrong magnetic fields
($B\gtrsim 10^{14}\mbox{ G}$)\footnote{However, there exists an apparently
low surface magnetic field magnetar \citep{reaetal10}.},
and their strong fields likely power the activity seen in these objects
(see \citealt{woodsthompson06,mereghetti08}, for review).
One notable property of magnetars is that their observed surface temperatures
$\Ts$ in quiescence are significantly higher than those of other NSs of a
similar age.  In fact, they are too high for NSs that cool passively, i.e.,
without an additional source of internal heat (accretion heating can be
excluded by, e.g., non-detections of binary companion or disk emission).
An interesting problem concerns the
heat generated from magnetic field decay, which has been proposed to be the
source for the high temperatures of magnetars
\citep{thompsonduncan96,heylkulkarni98,colpietal00,aguileraetal08b}.
This heat can strongly influence the time/age at which the core becomes
superfluid if heating/field decay occurs in the core
\citep{thompsonduncan96,arrasetal04,dallossoetal09}.
The problem is important since the presence of superfluid components has a
strong impact on magnetar interior dynamics, such as the mechanism for
producing glitches (see, e.g., \citealt{sauls89}),
fluid oscillations (see, e.g., \citealt{passamontiandersson11}),
magnetic field and rotational evolution
(see, e.g., \citealt{glampedakisandersson11}),
and magnetohydrodynamical equilibrium \citep{glampedakisetal12,landeretal12}.

We address this problem by conducting detailed calculations of the thermal
evolution of a NS with various prescriptions for an internal heat source
that can be associated with magnetic field decay.
We show that, regardless of the magnetic field strength and detailed
mechanism for field decay in the core, the heat generated by the decay is
insufficient to power the surface emission of magnetars.
Furthermore, by accounting for the effects of proton superconductivity on
field decay, we find that this core heating is not strong
enough to balance neutrino cooling and cannot delay the onset of core
neutron superfluidity for fields $\lesssim 10^{16}\mbox{ G}$
(onset at age $\lesssim\mbox{a few}\times 100\mbox{ yr}$).
Thus the cores of all currently known magnetars should be in a superfluid state.

We briefly describe past works and note key findings and assumptions
made in these works that we improve upon here.
\citet{arrasetal04} consider coupled magnetic field decay and thermal
evolution of magnetars.
The internal field and superfluid temperatures are assumed to be
$>10^{15}\mbox{ G}$ and $\Tcp=5\times 10^9\mbox{ K}$ and
$\Tcnt=(5-9)\times 10^8\mbox{ K}$, respectively.
\citet{arrasetal04} find that magnetic field decay can delay the
transition to core neutron superfluidity and maintain a relatively high
surface temperature to ages $\approx 10^3-10^5\mbox{ yr}$, depending on $\Tcnt$.
However, their calculation only considers volume-averaged quantities
and thus assumes that the NS interior is isothermal
(see Section~\ref{sec:nscooleq}),
which cannot be the case if there is a localized heating source such as
field decay in the crust or core due to the large but finite thermal
conductivity of NS matter.

\citet{kaminkeretal06} calculate the evolution of the temperature profile
[i.e., $T(\rho)$] by solving the energy balance and flux equations
[see eqs.~(\ref{eq:energybalancegr}) and (\ref{eq:heatfluxgr})].
They demonstrate that, in order to explain the observed high surface
temperatures, magnetars require a heat source and, most importantly,
this heat source (e.g., from field decay; \citealt{ponsetal09})
must be located in the outer crust; if the heat source is
located too deep in the NS interior (e.g., in the core), then neutrino
emission efficiently removes the heat locally, and the surface temperature
cannot be increased sufficiently to match the observed values.
Furthermore, the outer crust is thermally decoupled from the core so that
heating of the crust does not affect (neutrino) cooling of the core.
Because of this thermal decoupling, the results
and conclusions of \citet{kaminkeretal06} are not particularly sensitive to
the state of matter in the core, e.g., superfluidity of the core nucleons.
Though no quantitative results are shown, \citet{kaminkeretal06} state that
cooling calculations with the effects of inner crust superfluidity and neutrino
emission by Cooper pair formation produce different temperature profiles,
but these effects do not change the surface temperature.  In follow-up work,
\citet{kaminkeretal09} improve their calculations and examine the effects
of light-element accreted envelopes and anisotropic heat conduction due to
the magnetic field in the envelope and outer crust.
While accreted envelopes can give higher surface temperatures for the same
core temperatures, core heating is still unable to produce surface
temperatures that are high enough to explain magnetar observations.

\citet{dallossoetal09} criticize the work of \citet{kaminkeretal06},
arguing that the phenomenological heating model considered
is independent of magnetic field and decays on
the wrong timescale [see eq.~(\ref{eq:crustheat})] and that the correct
heating should depend inversely on temperature.
This last point implies that more heat is generated at lower temperatures
and can lead to an equilibrium between neutrino cooling and heating from
field decay (see also \citealt{thompsonduncan96}).
The equilibrium temperature (above $10^9\mbox{ K}$) can be maintained for
$\approx 10^4\mbox{yr}$.  The results of \citet{dallossoetal09}
suggest that the core of magnetars are not superfluid until
after this time, since the critical temperature for the onset of neutron
superfluidity is $(5-9)\times 10^8\mbox{ K}$ \citep{pageetal11,shterninetal11}.
However no cooling calculation is performed by \citet{dallossoetal09}.

Here we perform NS cooling simulations to determine the role of core heating
by magnetic field decay.
We do not examine in detail field evolution and heating in the crust
(see, e.g., \citealt{ponsetal09}).
Rather we use the phenomenological model of
\citet{kaminkeretal06,kaminkeretal09} to demonstrate the effect of
{\it crust heating} on the NS cooling behavior.
This will be adequate for our purposes since the goal here is to assess
the importance of {\it core heating} in magnetars.
As we will show, as long as the magnetic field decay time is longer than
the cooling time, the core heating we use is the maximum one can use.
While this maximal heating can delay the onset of neutron superfluidity
[on the timescale estimated by \citet{thompsonduncan96,dallossoetal09}],
it cannot reproduce the high surface temperatures seen for magnetars.
We arrive at this same conclusion using the prescription for heating
and field decay given in \citet{dallossoetal09}.
Furthermore, the above works neglect the effects of proton superconductivity;
when this is taken into account, there is no delay in superfluidity onset
for any reasonable core magnetic field strength.
In Section~\ref{sec:model}, we describe the thermal evolution equations
and input physics, including superfluid properties and prescriptions for
internal heat sources.
Section~\ref{sec:results} presents the results of our calculations.
We summarize our results and discuss their implications in
Section~\ref{sec:discuss}.

%%%%%%%%%%%%%%%%%%%%%%%%%%%%%%%%%%%%%%%%%%%%%%%%%%%%%%%%%%%%%%%%%%%
\section{Neutron Star Cooling Model} \label{sec:model}

%%%%%%%%%%%%%%%%%%%%%%%%%%%%%%%%%%%%%%%%%%%%%%%%%%%%%%%%%%%%%%%%%%%
\subsection{Core and Crust Composition} \label{sec:composition}

We use a stellar model based on the Akmal-Pandharipande-Ravenhall (APR)
equation of state (EOS)
\citep{akmaletal98,heiselberghjorthjensen99},
specifically APR I\footnote{The maximum NS mass for this EOS is
$1.923\,M_\odot$, which is below the highest measured mass of $1.97\,M_\odot$
\citep{demorestetal10}.  Nevertheless, APR is typical of EOSs that yield
higher maximum masses, and our conclusions do not depend strongly on the
specific EOS.}\citep{gusakovetal05}.
For a NS mass $M=1.4\,M_\odot$, the NS radius is $R=12.1\mbox{ km}$, and the
central density is $\rho_{\mathrm{c}}=9.47\times 10^{14}\mbox{ g cm$^{-3}$}$.
The crust composition\footnote{See http://relativity.livingreviews.org/Articles/lrr-2008-10/}
(in particular, charge number $Z$ and mass number $A$ of the ions)
as a function of density $\rho$ is determined by interpolating the values
taken from Table~VII of \citet{rusteretal06} for
$8.02\times 10^{6}\mbox{ g cm$^{-3}$}\le\rho
\le 4.27\times 10^{11}\mbox{ g cm$^{-3}$}$ and
Table~3 of \citet{negelevautherin73} for
$4.67\times 10^{11}\mbox{ g cm$^{-3}$}\le\rho
\le 7.94\times 10^{13}\mbox{ g cm$^{-3}$}$
and estimated from Figs.~6 and 9 of \citet{oyamatsu93} for
$\rho=(1,1.5)\times 10^{14}\mbox{ g cm$^{-3}$}$.
We note that the melting temperature $\Tmelt$ is given by
(see, e.g., \citealt{shapiroteukolsky83})
\be
\Tmelt=3.04\times 10^7\mbox{ K }(Z/26)^{5/3}(170/\Gamma_{\mathrm{m}})x,
 \label{eq:tmelt}
\ee
where $x=[(Z/A)(\rho/10^6\mbox{ g cm$^{-3}$})]^{1/3}$ and
$\Gamma_{\mathrm{m}}\approx 170$ is the value of the Coulomb parameter
$\Gammac$ [$=Z^2e^2/a\kb T$, where $a=(3/4\pi n_\mathrm{i})^{1/3}$ and
$n_\mathrm{i}$ is the ion number density] at which melting occurs.

%%%%%%%%%%%%%%%%%%%%%%%%%%%%%%%%%%%%%%%%%%%%%%%%%%%%%%%%%%%%%%%%%%%
\subsection{Equations for Neutron Star Cooling} \label{sec:nscooleq}

The evolution of the interior temperature $T(r,t)$ of an isolated NS is
determined by the relativistic equations of energy balance and heat flux,
i.e.,
\be
\frac{e^{-\Lambda-2\Phi}}{4\pi r^2}\frac{\partial}{\partial r}
 \lp e^{2\Phi}\Lr\rp = -e^{-\Phi}\heatcap\frac{\partial T}{\partial t}
 - \varepsilon_\nu + \Qheat \label{eq:energybalancegr}
\ee
\be
\frac{\Lr}{4\pi r^2} = - e^{-\Lambda-\Phi}\kdiff\frac{\partial}{\partial r}
 \lp e^\Phi T\rp, \label{eq:heatfluxgr}
\ee
where $\Lr$ is the luminosity at radius $r$, $\heatcap$ is the heat capacity,
$\varepsilon_\nu$ is the neutrino emissivity, $\Qheat$ is the
internal heating source, and $\kdiff$ is the thermal conductivity
(see, e.g., \citealt{thorne77,shapiroteukolsky83}).
The metric function $\Lambda$ is defined by the enclosed mass $m(r)$, i.e.,
$e^{\Lambda(r)}\equiv\lb 1-2Gm(r)/c^2r\rb^{-1/2}$,
$\Phi(r)$ is the metric function (gravitational potential in Newtonian limit)
that is obtained from $d\Phi/dr=-(dP/dr)/[\rho(1+P/\rho)]$,
and $e^{-\Lambda(R)}=e^{\Phi(R)}$ at the NS surface.
We define (see, e.g., \citealt{vanriper91})
\be
\tilde{T} \equiv e^{\Phi}T \qquad\mbox{and}\qquad
\tilde{L} \equiv e^{2\Phi}\Lr,
\ee
so that eqs.~(\ref{eq:energybalancegr}) and (\ref{eq:heatfluxgr}) become
\be
\frac{\partial\tilde{L}}{\partial r} =
 - 4\pi r^2 e^{\Lambda}\heatcap\frac{\partial\tilde{T}}{\partial t}
 - 4\pi r^2 e^{\Lambda+2\Phi}\lp\varepsilon_\nu-\Qheat\rp
 \label{eq:energybalancegr2}
\ee
\be
\tilde{L} = -4\pi r^2\kdiff e^{-\Lambda+\Phi}
 \frac{\partial\tilde{T}}{\partial r}, \label{eq:heatfluxgr2}
\ee
respectively.
Thus the inputs that determine how a NS cools are the
heat capacity $\heatcap$, neutrino emissivity $\varepsilon_\nu$,
thermal conductivity $\kdiff$, and heat source $\Qheat$.
The following sections discuss the physics that determine the four physical
parameters, $\heatcap$, $\varepsilon_\nu$, $\kdiff$, and $\Qheat$.
We note that, due to the high thermal conductivity, NSs without additional
heat sources ($\Qheat=0$) become essentially isothermal after
$\sim 10-100\mbox{ yr}$ \citep{lattimeretal94,gnedinetal01,yakovlevetal11}.
When isothermal, the thermal evolution is determined solely by
eq.~(\ref{eq:energybalancegr2}), which can be simplified by integrating
over the volume.
The evolution of the interior temperature is solved by a method similar to
that described in \citet{gnedinetal01}
(see also \citealt{yakovlevpethick04,pageetal06}, for review).

%%%%%%%%%%%%%%%%%%%%%%%%%%%%%%%%%%%%%%%%%%%%%%%%%%%%%%%%%%%%%%%%%%%
\subsection{Superfluid Properties} \label{sec:sf}

Superfluidity has two important effects on neutrino emission and NS cooling:
(1) suppression of heat capacities and emission mechanisms, like modified
Urca processes, that involve superfluid constituents and
(2) enhanced emission due to Cooper pairing of nucleons when the
temperature decreases just below the critical value
(see \citealt{yakovlevpethick04,pageetal06}, for review).

\begin{figure}
\begin{center}
\includegraphics[scale=0.4]{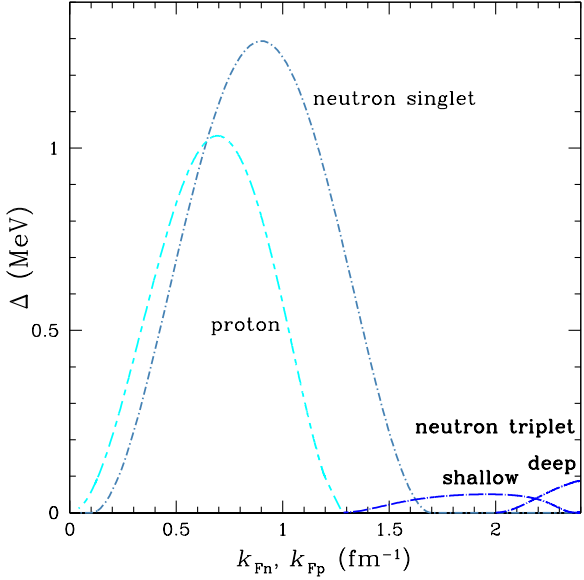}
\caption{
Superfluid gap energies (left: singlet; right: triplet) as a function of
Fermi wavenumber for neutrons $\kfn$ and protons $\kfp$.
The neutron singlet (dot-short-dashed) model is from \citet{wambachetal93},
the proton singlet (dashed) model is similar to \citet{chenetal93}, and
the neutron triplet (dot-long-dashed) model is either a shallow model
from \citet{elgaroyetal96} or a deep model similar to \citet{shterninetal11}.
}
\label{fig:tsfk}
\end{center}
\end{figure}

\begin{figure}
\begin{center}
\includegraphics[scale=0.4]{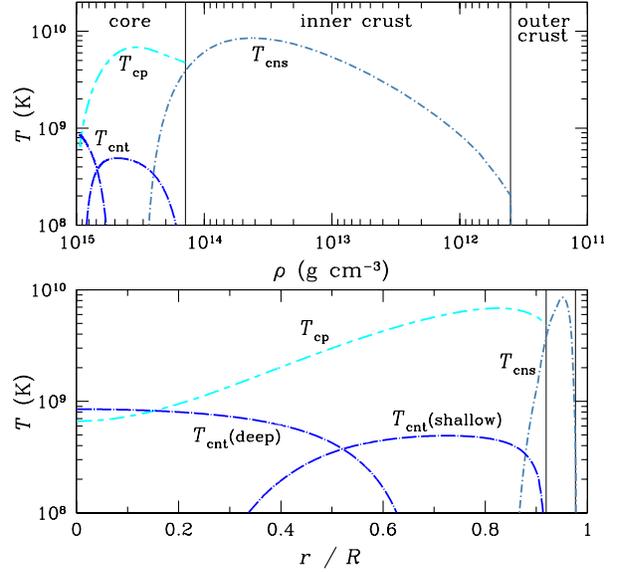}
\caption{
Superfluid critical temperatures as a function of density (top panel)
and normalized stellar radius (bottom panel).
Critical temperatures $\Tcns$ (dot-short-dashed), $\Tcnt$
(dot-long-dashed), and $\Tcp$ (dashed) are for neutron singlet,
neutron triplet, and proton singlet, respectively.
Neutron triplet pairing is taken to be described by either the shallow or
deep model (see text).
Vertical solid lines indicate the approximate boundaries between the
core and inner crust (at near nuclear saturation, i.e., $\rho\approx\densnuc/2$)
and inner and outer crusts (at neutron drip, i.e.,
$\rho\approx 4\times 10^{11}\mbox{ g cm$^{-3}$}$).
}
\label{fig:tsfr}
\end{center}
\end{figure}

The critical temperatures for superfluidity are approximately related
to the superfluid energy gap $\Delta$ by
$k\Tc\approx 0.5669\Delta$ for the singlet (isotropic pairing) gap and
$k\Tc\approx 0.8418\Delta$ for the maximum triplet (anisotropic pairing) gap.
We use the parametrization for the gap energy
\be
\Delta(\kf) = \Delta_0\frac{(\kf-k_0)^2}{(\kf-k_0)^2+k_1}
 \frac{(\kf-k_2)^2}{(\kf-k_2)^2+k_3},
\ee
where $\pf=\hbar\kf=\hbar(3\pi^2n_{\mathrm{x}})^{1/3}$ and $n_{\mathrm{x}}$
are the Fermi momentum and number density, respectively, for particle
${\mathrm{x}}$
and $\Delta_0$, $k_0$, $k_1$, $k_2$, and $k_3$ are fit parameters for
particular superfluid gap models given in \citet{anderssonetal05}
(see also \citealt{lombardoschulze01,kaminkeretal02}).
For singlet neutrons in the crust, we use model {\it a}, which represents
the results of \citet{wambachetal93}.
For singlet protons in the core, we use a model that is similar to 
\citet{chenetal93} (see also \citealt{pageetal04}).
We consider two cases for triplet neutron pairing in the core.
Either the pairing is the ``shallow'' model {\it l}, which represents the
results of \citet{elgaroyetal96}, or the pairing is a ``deep'' model.
The deep model is a $\Delta$ that produces a $\Tcnt(\rho)$ similar to what
is needed to fit the observed surface temperature of the neutron star in
Cassiopeia~A and neutron stars that are old and hot
\citep{gusakovetal04,gusakovetal05,shterninetal11}.
The parameters for the superfluid gap energies are given in Table~\ref{tab:sf},
and the gap models are illustrated in Figs.~\ref{fig:tsfk} and \ref{fig:tsfr}.

%-------------------------------------------
\begin{table}
\caption{Phenomenological parameters for superfluid gap energy \label{tab:sf}}
\begin{tabular}{ccccc}
\hline
 & & & shallow & deep \\
 & proton & neutron & neutron & neutron \\
 & singlet & singlet & triplet & triplet \\
\hline
$\Delta_0$ (MeV) & 120 & 68 & 0.068 & 0.15 \\
$k_0$ (fm$^{-1}$) & 0 & 0.1 & 1.28 & 2 \\
$k_1$ (fm$^{-2}$) & 9 & 4 & 0.1 & 0.1 \\
$k_2$ (fm$^{-1}$) & 1.3 & 1.7 & 2.37 & 3.1 \\
$k_3$ (fm$^{-2}$) & 1.8 & 4 & 0.02 & 0.02 \\
\hline
\end{tabular}
\end{table}
%-------------------------------------------

%-------------------------------------------
%-------------------------------------------

%-------------------------------------------
%-------------------------------------------

%%%%%%%%%%%%%%%%%%%%%%%%%%%%%%%%%%%%%%%%%%%%%%%%%%%%%%%%%%%%%%%%%%%
\subsection{Heat Capacity} \label{sec:heatcap}

In the core, the total heat capacity is the sum of partial heat capacities
due to neutrons, protons, electrons, and muons.
In the crust, the total heat capacity is the sum of partial heat capacities
due to free neutrons, ions, and electrons.
The partial heat capacity for a strongly degenerate fermion particle species
${\mathrm{x}}$ is
\be
\heatcapj = \frac{\meff\pf\kb^2T}{3\hbar^3},
\ee
where $\meff$ is the effective mass.
We assume $\meff=0.7$ for neutrons and protons and
take $\meff=(m_{\mathrm{x}}^2c^4+\pf^2c^2)^{1/2}$ for electrons and muons.
A reduction of the heat capacity due to nucleon superfluidity can be
taken into account by using
\be
\heatcapsfj = \redheatcapj\heatcapj
\ee
where the reduction factor $\redheatcapj$ only depends on the ratio
$T/\Tcj$ and whether the superfluid is of singlet or triplet type
\citep{levenfishyakovlev94,yakovlevetal99}.
The ion heat capacity is \citep{vanriper91}
\be
\heatcap_{\mathrm{ion}} = (3/2)\kb\times \left\{ \begin{array}{ll}
1 & \mbox{if } \Gammac\le 1 \\
2f_{\mathrm{D}}(T/\Tdebye) & \mbox{if } 1<\Gammac\le 150 \\
1+\frac{\log\Gammac}{\log 150} & \mbox{if } \Gammac>150 \end{array} \right.,
\ee
where
\be
f_{\mathrm{D}}(T/\Tdebye) = \left\{ \begin{array}{ll}
0.8\pi^4(T/\Tdebye)^3 & \mbox{if } T/\Tdebye\le0.15 \\
1-0.05(\Tdebye/T)^2 & \mbox{if } T/\Tdebye\ge 4 \\
1.70(T/\Tdebye)+0.0083 & \mbox{otherwise} \end{array} \right.
\ee
and the Debye temperature is
\be
\Tdebye=3.48\times 10^6\mbox{ K }(Z/A)(\rho/10^6\mbox{ g cm$^{-3}$})^{1/2}.
\ee

%%%%%%%%%%%%%%%%%%%%%%%%%%%%%%%%%%%%%%%%%%%%%%%%%%%%%%%%%%%%%%%%%%%
\subsection{Neutrino Emissivity} \label{sec:emissivity}

For the NS core, we calculate neutrino emission due to the neutron and proton
branches of the modified Urca process and neutron-neutron, neutron-proton,
and proton-proton bremsstrahlung using emissivities from
\citet{yakovlevetal99,pageetal04}.
When neutrons and/or protons are superfluid, we account for suppression
of the above processes \citep{levenfishyakovlev94b,yakovlevetal99}
and neutrino emission due to Cooper pairing of the superfluid component
\citep{yakovlevetal99,pageetal09}.

For the APR I EOS (see Section~\ref{sec:composition}), neutrino emission by
the direct Urca process occurs when $M>1.829\,M_\odot$
(central densities above $1.680\times 10^{15}\mbox{ g cm$^{-3}$}$;
\citealt{gusakovetal05});
for this present work
in which we consider only $M=1.4\,M_\odot$, we will neglect this process
(as well as other `fast' cooling processes;
see \citealt{yakovlevpethick04,pageetal06}).
As shown by \citet{kaminkeretal06}, cooling by direct Urca has no effect
on the surface temperature because of thermal decoupling between the core
and outer crust.
In addition, since direct Urca is more effective at cooling the core,
it would speed up the onset of superfluidity. Hence its inclusion would
make our conclusions even more robust.

In the crust, we account for electron-nucleon, neutron-neutron, and
neutron-nucleon bremsstrahlung, plasmon decay, and e$^-$-e$^+$ pair
annihilation using emissivities from \citet{yakovlevetal99a,yakovlevetal01}
When neutrons are superfluid (in the singlet state), we account for the
suppression of neutron-neutron and neutron-nucleon bremsstrahlung
\citep{yakovlevetal99,yakovlevetal01},
as well as including neutrino emission due to neutron Cooper pairing
\citep{yakovlevetal99,pageetal09}.
We neglect neutrino synchrotron emission in a magnetic field since this
additional emission would cause even more rapid cooling of the crust
\citep{yakovlevetal01}.

%%%%%%%%%%%%%%%%%%%%%%%%%%%%%%%%%%%%%%%%%%%%%%%%%%%%%%%%%%%%%%%%%%%
\subsection{Thermal Conductivity} \label{sec:thermalcond}

For the core, we sum thermal conductivities due to neutrons, electrons,
and muons \citep{flowersitoh79,flowersitoh81}.
We use the results of \citet{baikoetal01} to calculate the neutron thermal
conductivity, accounting for neutron-neutron and neutron-proton collisions and
superfluid suppression.
We use the results of \citet{shterninyakovlev07} to calculate the electron
and muon thermal conductivities and account for superfluid suppression.
For the crust,
we use {\small CONDUCT08}\footnote{http://www.ioffe.ru/astro/conduct/},
which implements the latest advancements in calculating thermal
conductivities \citep{potekhinetal99,cassisietal07,chugunovhaensel07}.
We assume no contribution by impurity scattering, which only becomes
important at high densities and low temperatures.

At high magnetic fields and low temperatures, electron motion is strongly
influenced by the magnetic field (see, e.g., \citealt{yakovlevkaminker94}).
As a result, the electrical and thermal conductivities become anisotropic,
depending on whether electrons move parallel or transverse to the direction
of the field \citep{potekhin99}.
However, electron magnetization and anisotropic heat conduction are weaker
at higher temperatures (see \citealt{lai01}, for review).
In particular, magnetic field effects are minimal when the field is in the
non- or weakly-quantizing regime, i.e., when $T>T_\mathrm{B}$ and
$\rho\gg\densmag$, where
\be
\Tmag = 1.34\times 10^{10}\mbox{ K }(B/10^{14}\mbox{ G})(1+x^2)^{-1/2},
 \label{eq:tmag}
\ee
$x$ is defined below eq.~(\ref{eq:tmelt}), and
$\densmag=7.046\times 10^{6}\mbox{ g cm$^{-3}$}(A/Z)(B/10^{14}\mbox{ G})^{3/2}$.
Since we are primarily concerned with effects in the high density core of
young NSs with high interior temperatures and the core is effectively
thermally decoupled from the crust \citep{kaminkeretal06,kaminkeretal09},
we neglect magnetic field effects on the conductivities below the heat blanket
(see also Section~\ref{sec:envelope}).
In fact, recent works show that the effects of superfluid phonons
\citep{aguileraetal09} and toroidal magnetic fields \citep{ponsetal09}
in the crust can very effectively smooth out temperature variations
induced by anisotropic heat conduction.

%%%%%%%%%%%%%%%%%%%%%%%%%%%%%%%%%%%%%%%%%%%%%%%%%%%%%%%%%%%%%%%%%%%
\subsection{Envelope} \label{sec:envelope}

The outer layers (envelope) of the NS crust serve as a heat blanket, and
there can exist a large temperature gradient between the bottom of the
envelope (at $\densb$ and $\Tb$) and the surface.
We assume $\densb=10^{10}\mbox{ g cm$^{-3}$}$
\citep[see, e..g,][]{gudmundssonetal82,yakovlevpethick04}.
\citet{changbildsten04} (see also \citealt{changetal10}) show that, at
the high temperatures present in magnetars, nuclear burning very rapidly
removes any surface hydrogen;
nevertheless, for generality, we consider surfaces composed of either
iron or light-elements.
For iron, the relationship between the temperature at the bottom of the
heat-blanketing envelope $\Tb$ and the effective temperature of the
photosphere $\Ts$ is \citep{gudmundssonetal82}
\be
\Ts = 8.7\times 10^5\mbox{ K }\sgravonefour^{1/4}
 \lp\Tb/10^8\mbox{ K}\rp^{11/20}, \label{eq:tsiron}
\ee
where $\sgravonefour=\sgrav/10^{14}\mbox{ cm s$^{-2}$}$ and
$\sgrav=(1-2GM/c^2R)^{-1/2}GM/R^2$ is the surface gravity
($\sgravonefour=1.55$ for our assumed NS model).
Non-magnetized envelopes composed of light elements (H or He) have been
considered in \citet{potekhinetal97}
[\citet{kaminkeretal09} find a negligible difference in the results using
either H or He at $\Tb>10^8\mbox{ K}$; see also \citealt{yakovlevetal11},
for carbon envelopes]. Due to the higher thermal conductivity
($\propto Z^{-1}$) of light elements, the difference between $\Ts$ and $\Tb$
is larger for a light-element envelope compared to an iron envelope,
i.e., $\Ts$ is higher for a given $\Tb$ and $\Tb$ is lower for
a given $\Ts$.
For a fully-accreted light element envelope \citep{potekhinetal97},
\be
\Ts = 1.43\times 10^6\mbox{ K }\sgravonefour^{1/4}
 \lp\Tb/10^8\mbox{ K}\rp^{17/28}. \label{eq:tsacc}
\ee
For clarity, we show the above relations
[eqs.~(\ref{eq:tsiron}) and (\ref{eq:tsacc})];
however, we use the more accurate expressions given in \citet{potekhinetal03}
in our calculations.

Magnetized envelopes produce higher (lower) $\Ts$ for a given $\Tb$
due to enhanced (reduced) thermal conduction along (across) the magnetic
field, and the effect is stronger for accreted envelopes
\citep{potekhinyakovlev01,potekhinetal03}.
\citet{kaminkeretal06,kaminkeretal09} include the effects of
magnetic fields ($\sim 10^{14}-10^{16}\mbox{ G}$)
in the envelope and inner and outer crusts
and find that, though the temperature profile in the crust is modified,
the average surface temperature is only weakly affected by anisotropic heat
conduction.
We mostly neglect magnetized envelopes since their effects would not
significantly affect our conclusions (see Section~\ref{sec:obs}).

%%%%%%%%%%%%%%%%%%%%%%%%%%%%%%%%%%%%%%%%%%%%%%%%%%%%%%%%%%%%%%%%%%%
\subsection{Heat Source} \label{sec:heatsource}

We consider several prescriptions for an internal heat source.
For heating in the crust, we use a model similar to the phenomenological
one of \citet{kaminkeretal06,kaminkeretal09}, i.e.,
\be
\Qheat = \Qheati \exp\{-[(\rho-\densheat)/\Deltaheat]^2\} \exp(-t/\tauheat);
\label{eq:crustheat}
\ee
the exact form of heating is unimportant, as demonstrated by
\citet{kaminkeretal09}.
We take $\tauheat=10^4\mbox{ yr}$,
$\Qheati=3\times 10^{20}\mbox{ ergs cm$^{-3}$ s$^{-1}$}$,
$\densheat=6\times 10^{10}\mbox{ g cm$^{-3}$}$, and
$\Deltaheat=2\times 10^{10}\mbox{ g cm$^{-3}$}$.
The total heat luminosity
$\Lheat=\int\Qheat e^{2\Phi}\,4\pi r^2dr\approx 10^{37}\mbox{ ergs s$^{-1}$}$
at $t\ll\tauheat$.
Note that \citet{kaminkeretal06,kaminkeretal09} use a heating decay
timescale $\tauheat=5\times 10^4\mbox{ yr}$;
the exact value is unimportant for our purposes, as long as it is
approximately the age of the oldest (high surface temperature) magnetar,
since $\tauheat$ is approximately the time during which a high surface
temperature can be maintained by this crust heating
(see Section~\ref{sec:obs}).

In the core, heating is supplied by magnetic field decay due to ambipolar
diffusion \citep{goldreichreisenegger92}. We estimate a heating rate
\ba
\Qheat &\approx& \frac{B}{4\pi}\frac{dB}{dt} \approx \frac{B^2}{4\pi\tamb}
 = 1\times 10^{20}\mbox{ ergs cm$^{-3}$ s$^{-1}$}
 \nonumber\\ && \times
 \lp\frac{\rho}{\densnuc}\rp^{-2/3}\lp\frac{T}{10^{9}\mbox{ K}}\rp^{-2}
 \lp\frac{B}{10^{16}\mbox{ G}}\rp^{4}, \label{eq:coreheat}
\ea
where the ambipolar diffusion timescale $\tamb$ is given by
\citep{goldreichreisenegger92}
\ba
\tamb &\sim& 2.5\times 10^3\mbox{ yr }(L/1\mbox{ km})^2(\rho/\densnuc)^{2/3}
 \nonumber\\ && \times
 (T/10^9\mbox{ K})^2(B/10^{16}\mbox{ G})^{-2} \label{eq:tambnonsf}
\ea
and we neglect here the proton fraction dependence
and assume the field decay lengthscale $L$ $(=1\mbox{ km})$ is
constant for simplicity. Heating is taken to occur throughout the core and
is thus very strong.
Note that eq.~(\ref{eq:coreheat}) is identical to what is considered by
\citet{dallossoetal09}, except for a small difference 
in the numerical coefficient.
It is also worth noting that $\tamb$ is the timescale of the ``solenoidal''
mode.  This mode is controlled by collisions between the different particle
species in NS matter and operates at relatively high temperatures
($\gtrsim\mbox{a few}\times 10^8\mbox{ K}$).  Superfluidity can significantly
modify $\tamb$ and thus the heating rate given by eq.~(\ref{eq:coreheat}).
This is discussed in more detail in Section~\ref{sec:onset}.

It is not well-understood how magnetic field decays (if at all) in the core
at times $t\lesssim 10^4\mbox{ yr}$
(see, e.g., \citealt{haenseletal90,goldreichreisenegger92,thompsonduncan96};
\citealt{glampedakisetal11}).
Therefore, to describe the decay of the core magnetic field, we use the
simple formula
\be
B(t)=\frac{B_0}{1+t/\tdecay}, \label{eq:magevol}
\ee
where $\tdecay$ is the field decay timescale,
which we choose to be $10^4\mbox{ yr}$.
Though eq.~(\ref{eq:magevol}) has no physical basis, it has a form
similar to ones considered in the literature, usually with the denominator
(i.e., time-dependence) to some low-order power (see, e.g.,
\citealt{thompsonduncan96,colpietal00,dallossoetal09}).\footnote{\citet{thompsonduncan96}
has a $1/14$-th power of the denominator and a decay timescale proportional
to $\tamb$, while \citet{dallossoetal09} has a $5/6$-th power of the
denominator and a decay timescale given by
$1.03\times 10^4\mbox{ yr }(B/10^{16}\mbox{ G})^{-6/5}
(\rho/10^{15}\mbox{ g cm$^{-3}$})^{2/5}$.
The different evolution laws reflect the different ambipolar diffusion modes
(solenoidal vs non-solenoidal) considered by these authors.
Note that these decay timescales are shorter for higher magnetic fields.
Incidentally, numerical simulations of field evolution in the crust yield
eq.~(\ref{eq:magevol}), with $\tdecay$ given by the timescale for Hall drift
at times much shorter than the Ohmic decay timescale
\citep{aguileraetal08,aguileraetal08b}.}
However, recall that our aim is to obtain a {\it maximum} core heating
rate, in order to test the extent to which core heating can produce the
observed (high) surface temperature of magnetars and delay the onset of
superfluidity.
It is clear that eq.~(\ref{eq:coreheat}) is maximal before significant
field decay, i.e., $B=B_0$ at $t\ll\tdecay$.
[At $t=10^3\mbox{ yr}$, we find
$\Lheat\sim 2\times 10^{38}\mbox{ ergs s$^{-1}$}(B/10^{16}\mbox{ G})^4$.]
Therefore, as long as $\tdecay$ is longer than the thermal diffusion
timescale, this core heating will have a maximum effect on the surface
temperature.
We also conduct simulations utilizing the heating rate and
field decay formulae from \citet{dallossoetal09} and find very similar
results at the ages of interest here ($<10^5\mbox{ yr}$).

%%%%%%%%%%%%%%%%%%%%%%%%%%%%%%%%%%%%%%%%%%%%%%%%%%%%%%%%%%%%%%%%%%%
\section{Results} \label{sec:results}

We assume a constant $Te^\Phi=10^{10}\mbox{ K}$ at $t=0$.
For clarity, the superfluid results shown in all subsequent figures use
the shallow neutron triplet model (see Section~\ref{sec:sf});
the superfluid results using the deep neutron triplet model are
qualitatively similar in the age range considered here
($<10^5\mbox{ yr}$).
Also we assume an iron envelope, unless noted otherwise.
For models which include core heating,
we primarily assume an initial magnetic field $B_0=10^{16}\mbox{ G}$,
though we also consider lower, more realistic initial field strengths,
as well as extreme fields.

%%%%%%%%%%%%%%%%%%%%%%%%%%%%%%%%%%%%%%%%%%%%%%%%%%%%%%%%%%%%%%%%%%%
\subsection{Cooling with no superfluid nor heating} \label{sec:nonsf}

\begin{figure}
\begin{center}
\includegraphics[scale=0.4]{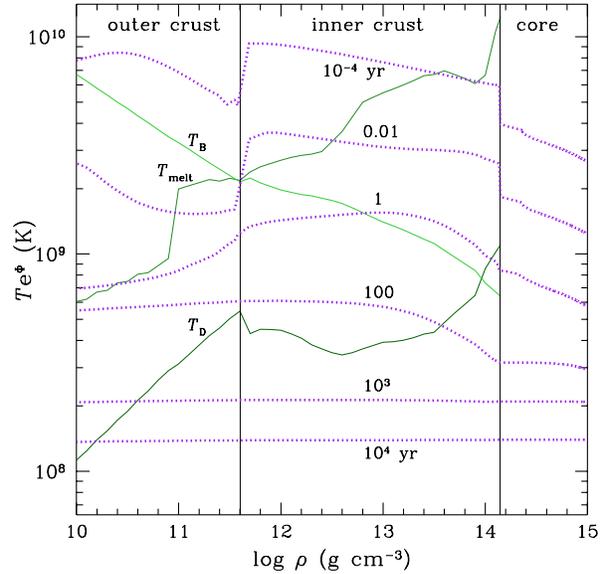}
\caption{
Temperature profiles for cooling model with no superfluidity
and no heating (dotted lines).
The six profiles are at ages $t=10^{-4}$ (top), 0.01, 1, 100, $10^3$, and
$10^4\mbox{ yr}$ (bottom).
Thin solid lines denote the melting temperature $\Tmelt$, Debye temperature
$\Tdebye$, and $\Tmag$, where $\Tmag$ is calculated assuming
$B=10^{15}\mbox{ G}$ [see eq.~(\ref{eq:tmag})].
Vertical solid lines indicate boundaries between core and inner crust and
inner and outer crusts.
}
\label{fig:tempprof}
\end{center}
\end{figure}

Figure~\ref{fig:tempprof} shows the temperature as a function of density
at different ages for a NS cooling model that neglects superfluidity and
has no additional sources of internal heating ($\Qheat=0$).
These temperature profiles are very similar to those shown in
\citet{gnedinetal01}.
At very early times, the core cools more rapidly than the crust via
stronger neutrino emission, so that the crust is generally at higher
temperatures.
A cooling wave travels from the core to the surface, bringing the NS to
a relaxed, isothermal state.  Depending on the properties of the crust,
the relaxation time is $\sim 10-100\mbox{ yr}$
\citep{lattimeretal94,gnedinetal01,yakovlevetal11}.
Formation of the inner and outer crusts begins at
$\sim 1\mbox{ hr}$ and $\sim 1\mbox{ day}$, respectively,
and is mostly complete after
$\sim 1\mbox{ month}$ and $\sim 1\mbox{ yr}$, respectively
(see also \citealt{aguileraetal08}).
The temperature profiles demonstrate the need to account for magnetic
field effects in more accurate models of the crust and envelope when
$T\lesssim\Tmag$ and temperature gradients are significant
(see Sections~\ref{sec:thermalcond} and \ref{sec:envelope}).

%-------------------------------------------
%-------------------------------------------

%%%%%%%%%%%%%%%%%%%%%%%%%%%%%%%%%%%%%%%%%%%%%%%%%%%%%%%%%%%%%%%%%%%
\subsection{Cooling with {\it np} superfluid and no heating} \label{sec:sfcool}

\begin{figure}
\begin{center}
\includegraphics[scale=0.4]{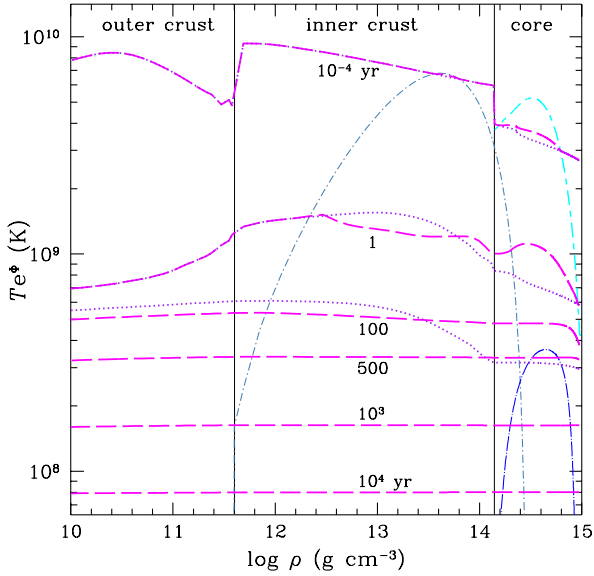}
\caption{
Temperature profiles for cooling model with superfluidity and no heating
(long-dashed lines).
The six profiles are at ages $t=10^{-4}$ (top), 1, 100, 500, $10^3$, and
$10^4\mbox{ yr}$ (bottom).
Also plotted for comparison are profiles (dotted lines) for the model with
no superfluidity at ages $t=10^{-4}$, 1, and $100\mbox{ yr}$
(see Fig.~\ref{fig:tempprof}).
Critical temperatures for neutron singlet (dot-short-dashed),
neutron triplet (dot-long-dashed), and proton singlet (short-long-dashed)
are shown.
Vertical solid lines indicate boundaries between core and inner crust and
inner and outer crusts.
}
\label{fig:tempprofsf}
\end{center}
\end{figure}

Figure~\ref{fig:tempprofsf} shows $T(\rho)$ at different ages for a cooling
model that neglects heating but includes superfluidity of neutrons and protons,
with critical temperatures also shown.
When superfluidity is taken into account, we see the impact of the
two effects mentioned in Section~\ref{sec:sf}:
Slower cooling in the core after protons become superconducting
and faster cooling after neutrons become
superfluid due to neutrino emission from Cooper pair formation.
The latter is strongest in regions near the critical temperature.
Proton superconductivity occurs at $\sim 1\mbox{ min}$,
and much of the core is superconducting after $\sim 1\mbox{ yr}$.
Core neutrons start becoming superfluid at around
$\mbox{a few}\times 100\mbox{ yr}$.

%-------------------------------------------
%-------------------------------------------

%%%%%%%%%%%%%%%%%%%%%%%%%%%%%%%%%%%%%%%%%%%%%%%%%%%%%%%%%%%%%%%%%%%
\subsection{Cooling with superfluid and crust heating} \label{sec:crust}

\begin{figure}
\begin{center}
\includegraphics[scale=0.4]{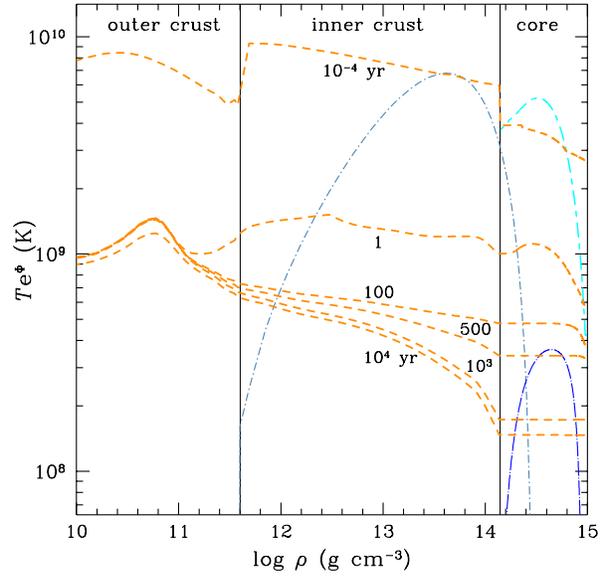}
\caption{
Temperature profiles for cooling model with superfluidity, crust heating,
and heating timescale $\tauheat=10^4\mbox{ yr}$ (short-dashed lines).
The six profiles are at ages $t=10^{-4}$ (top), 1, 100, 500, $10^3$, and
$10^4\mbox{ yr}$ (bottom).
Critical temperatures for neutron singlet (dot-short-dashed),
neutron triplet (dot-long-dashed), and proton singlet (short-long-dashed)
are shown.
Vertical solid lines indicate boundaries between core and inner crust and
inner and outer crusts.
}
\label{fig:tempprofcrust}
\end{center}
\end{figure}

Figure~\ref{fig:tempprofcrust} shows temperature profiles for a cooling model
that includes superfluidity and crust heating [see eq.~(\ref{eq:crustheat})].
The profiles with crust heating are similar to those shown in
\citet{kaminkeretal06,kaminkeretal09}.
In particular, it is clear that an additional heat source in the outer
crust can very effectively maintain a high temperature near the NS surface
[with redshifted surface temperature $\Ts^\infty>3\times 10^6\mbox{ K}$,
where $\Ts^\infty=\Ts(1-2GM/c^2R)^{1/2}$, for times longer than $\tauheat$
($=10^4\mbox{ yr}$)],
but this strong heating in the crust does not prevent the core from
cooling rapidly.
Due to thermal decoupling between the outer crust and core,
the core temperature drops below the critical temperature for neutron
triplet superfluidity at $\mbox{a few}\times 100\mbox{ yr}$,
whether or not there is crust heating.

%-------------------------------------------
%-------------------------------------------

%%%%%%%%%%%%%%%%%%%%%%%%%%%%%%%%%%%%%%%%%%%%%%%%%%%%%%%%%%%%%%%%%%%
\subsection{Cooling with superfluid and normal core heating} \label{sec:core}

\begin{figure}
\begin{center}
\includegraphics[scale=0.4]{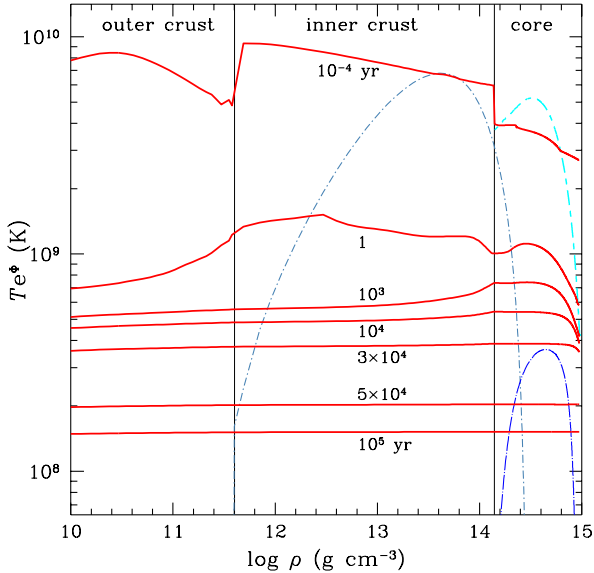}
\caption{
Temperature profiles for cooling model with superfluidity, normal core heating,
initial magnetic field $B_0=10^{16}\mbox{ G}$, and field decay timescale
$\tdecay=10^4\mbox{ yr}$ (solid lines).
The seven profiles are at ages $t=10^{-4}$ (top), 1, $10^3$, $10^4\mbox{ yr}$,
$3\times 10^4$, $5\times 10^4$, and $10^5\mbox{ yr}$ (bottom).
Critical temperatures for neutron singlet (dot-short-dashed),
neutron triplet (dot-long-dashed), and proton singlet (short-long-dashed)
are shown.
Vertical solid lines indicate boundaries between core and inner crust and
inner and outer crusts.
}
\label{fig:tempprofcore}
\end{center}
\end{figure}

Figure~\ref{fig:tempprofcore} shows temperature profiles for a model that
includes superfluidity and core heating due to magnetic field decay with
an initial field $B_0=10^{16}\mbox{ G}$ and $\tdecay=10^4\mbox{ yr}$
[see eqs.~(\ref{eq:coreheat}) and (\ref{eq:magevol})].
We see that, with such strong heating, the core temperature stays above the
critical temperature $\Tcnt$ for $t\gg 100\mbox{ yr}$.
However, the extra heat created in the core from field decay is efficiently
removed by core neutrino emission processes, with their strong temperature
dependencies \citep{yakovlevpethick04,pageetal06};
while the temperatures in the crust are higher than those from models
without any heating, they are significantly lower than those from models
with crust heating.

%-------------------------------------------
%-------------------------------------------

%%%%%%%%%%%%%%%%%%%%%%%%%%%%%%%%%%%%%%%%%%%%%%%%%%%%%%%%%%%%%%%%%%%
\subsection{Comparison with magnetar surface temperatures} \label{sec:obs}

\begin{figure}
\begin{center}
\includegraphics[scale=0.4]{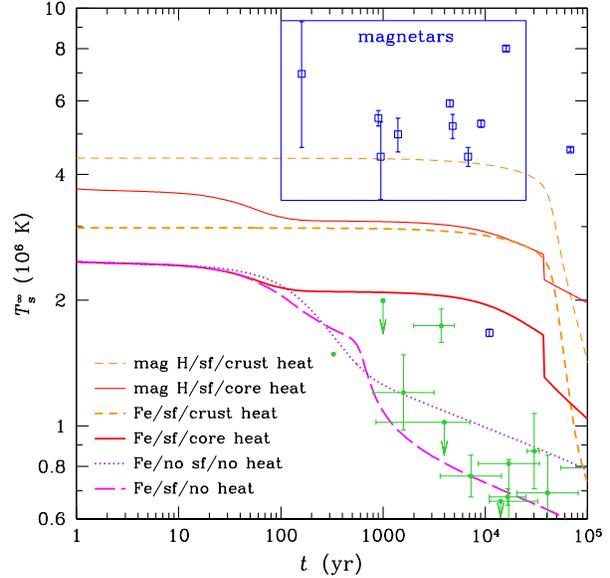}
\caption{
Redshifted surface temperature evolution for models with
no superfluid (dotted line),
superfluid and no heating (long-dashed line),
superfluid and crust heating for heating timescale
$\tauheat=10^4\mbox{ yr}$ (short-dashed lines), and
superfluid and core heating for initial magnetic field
$B_0=10^{16}\mbox{ G}$ and field decay timescale $\tdecay=10^4\mbox{ yr}$
(solid lines);
the thin upper lines have fully accreted light-element envelopes
(with a $10^{15}\mbox{ G}$ radial surface magnetic field),
while the thick lower lines have iron envelopes.
Data points are magnetars and other neutron stars taken from the McGill
SGR/AXP Online Catalog and those listed in
\citet{chevalier05,yakovlevetal08,hoheinke09,kaminkeretal09}, respectively.
}
\label{fig:timets}
\end{center}
\end{figure}

Figure~\ref{fig:timets} shows the evolution of redshifted surface
temperatures $\Ts^\infty$ for the cooling models plotted in
Figs.~\ref{fig:tempprof} -- \ref{fig:tempprofcore}.
For models with crust or core heating, we also show cooling curves with
fully accreted light element envelopes
and a surface magnetic field $10^{15}\mbox{ G}$ that is directed along the
radial direction, which produce higher surface temperatures for the same
outer crust temperature (\citealt{potekhinetal03};
see also Section~\ref{sec:envelope}).
The cooling curves are compared to the estimated ages and measured surface
temperatures of magnetars\footnote{Data are taken from the McGill SGR/AXP
Online Catalog at http://www.physics.mcgill.ca/$\sim$pulsar/magnetar/main.html.
Magnetar ages and surface temperatures are not well-determined due to
additional systematic uncertainties that are large
(see, e.g., \citealt{kaminkeretal06,kaminkeretal09}, for details).
\citet{kaminkeretal09} consider the more reliable surface luminosity.
For simplicity, we consider the ``magnetar box'' illustrated in
Fig.~\ref{fig:timets}.} and other neutron stars
(see \citealt{chevalier05,yakovlevetal08,hoheinke09,kaminkeretal09},
and references therein).
Our results clearly indicate that heating in the (outer) crust can produce
surface temperatures which match the high magnetar surface temperatures; this
is in agreement with the findings of \citet{kaminkeretal06,kaminkeretal09}.
On the other hand, even using a more realistic core heating prescription
that includes magnetic field strength and temperature dependencies
(as suggested by \citealt{dallossoetal09}),
heating of the core by magnetic field decay produces $\Ts^\infty$ that
are too low to explain the observed temperatures of magnetars.
Only for extreme ($B_0>10^{16}\mbox{ G}$)
and long-lived ($\tdecay\gtrsim 10^3\mbox{ yr}$) magnetic fields
and a magnetic, fully accreted light element envelope (which is unlikely due to
diffusive nuclear burning in the hot surface layers; \citealt{changbildsten04})
could a surface temperature at the low end of magnetar temperatures
($\Ts^\infty\sim 3\times 10^6\mbox{ K}$) be produced.
Therefore core heating alone cannot explain magnetar surface temperatures.

%-------------------------------------------
%-------------------------------------------

%%%%%%%%%%%%%%%%%%%%%%%%%%%%%%%%%%%%%%%%%%%%%%%%%%%%%%%%%%%%%%%%%%%
\subsection{Onset of core neutron superfluidity with proton superconductivity}
 \label{sec:onset}

From Figs.~\ref{fig:tempprofsf} -- \ref{fig:tempprofcore},
we see that the onset of core neutron superfluidity can be delayed from age
$\mbox{a few}\times 100\mbox{ yr}$ when the initial core field
$B_0\lesssim\mbox{a few}\times 10^{15}\mbox{ G}$ to,
e.g., $\approx 10^4\mbox{ yr}$ when $B_0=5\times 10^{15}\mbox{ G}$
(and $\tdecay=10^4\mbox{ yr}$)
or $3\times 10^4\mbox{ yr}$ when $B_0=8\times 10^{16}\mbox{ G}$
(and $\tdecay=10^3\mbox{ yr}$).
This is based on our simple model of core heating, especially the assumption
that field decay occurs on the timescale given by eq.~(\ref{eq:tambnonsf}).
For the deep neutron triplet model (see Sec.~\ref{sec:sf}), onset occurs
earlier since the inner core cools faster and the maximum $\Tcnt$ is higher
than in the shallow model (see Fig.~\ref{fig:tsfr}).

The decay timescale given by eq.~(\ref{eq:tambnonsf}) is the timescale for
the solenoidal mode of ambipolar diffusion of normal
(non-superconducting) protons \citep{goldreichreisenegger92,thompsonduncan96}.
At lower temperatures (later times), there is a transition of ambipolar
diffusion from the solenoidal to the non-solenoidal mode. This latter mode
is driven by departure from beta chemical equilibrium and has a timescale
\citep{goldreichreisenegger92}
\be
\tambir\sim 240\mbox{ yr }(\rho/\densnuc)^{4/3}(B/10^{16}\mbox{ G})^{-2}
 (T/10^9\mbox{ K})^{-6}, \label{eq:tambirnosf}
\ee
and the transition occurs at temperature
\be
\Ttr=7\times 10^8\mbox{ K }(\rho/\densnuc)^{1/12}. \label{eq:temptrnonsf}
\ee

\begin{figure}
\begin{center}
\includegraphics[scale=0.4]{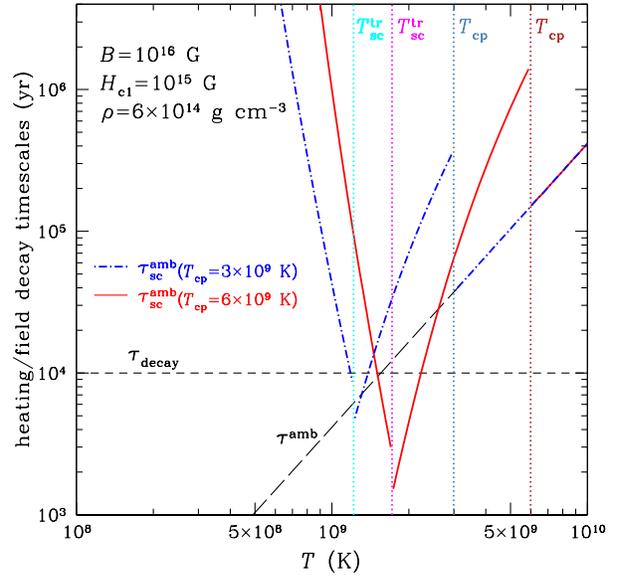}
\caption{
Core heating and field decay timescales used in this work as a function of
temperature for typical parameter values.
Long-dashed line is (solenoidal) ambipolar diffusion timescale $\tamb$ in
normal matter [see eq.~(\ref{eq:tambnonsf})],
dot-dashed and solid lines are ambipolar diffusion timescales $\ttamb$
which account for proton superconductivity at $\Tcp=3\times 10^9\mbox{ K}$
and $6\times 10^9\mbox{ K}$, respectively [see eq.~(\ref{eq:ttamb})],
and short-dashed line is the field decay timescale $\tdecay$
[$=10^4\mbox{ yr}$; see eq.~(\ref{eq:magevol})].
Vertical dotted lines denote the critical temperatures for proton
superconductivity $\Tcp$
and temperatures for transition between solenoidal and non-solenoidal
modes of ambipolar diffusion $\Ttrsc$ [see eq.~(\ref{eq:temptrsf})].
}
\label{fig:tdecay}
\end{center}
\end{figure}

Since $\Tcp\sim (2-3)\times 10^9\mbox{ K}$ \citep{pageetal11,shterninetal11},
proton superconductivity occurs very early.
When protons are superconducting (and neutrons are normal), processes
involving protons are suppressed
(see \citealt{glampedakisetal11}, for more detailed calculation).
In particular, the beta equilibrium reaction rate is reduced by a factor
$R_{p{\mathrm{A}}}$, which depends only on the ratio $T/\Tcp$
\citep{haenseletal01},
and the solenoidal and non-solenoidal ambipolar diffusion timescales are
given by
\be
\tambssc=(\tilde{R}_{p}B/H_{\mathrm{c1}})\tamb \label{eq:tambssc}
\ee
\be
\tambirsc=(B/H_{\mathrm{c1}}R_{p{\mathrm{A}}})\tambir, \label{eq:tambirsc}
\ee
respectively, where $H_{\mathrm{c1}}$ ($\approx 10^{15}\mbox{ G}$) is the
first critical magnetic field,
$\tilde{R}_{p}$ [$=R_{p}+(\meffe/m_e)(D_{en}/D_{pn})$] is the reduction
factor modified to account for electron-neutron collisions, $R_{p}$ is the
reduction factor of proton-neutron collisions given in \citet{baikoetal01},
and $D_{en}=10^{14}\mbox{ s$^{-1}$ }(\rho/\densnuc)^{-2/3}(T/10^9\mbox{ K})^2$
and $D_{pn}=4.8\times 10^{18}\mbox{ s$^{-1}$ }(\rho/\densnuc)^{-1/3}
(T/10^9\mbox{ K})^2$ are the inverse relaxation timescales for $e$-$n$
and $p$-$n$ collisions, respectively \citep{yakovlevshalybkov90}.
The transition temperature also increases, i.e.,
\be
\Ttrsc=(R_{p{\mathrm{A}}}\tilde{R}_{p})^{-1/8}\Ttr.  \label{eq:temptrsf}
\ee
Thus the heating rate in the core due to magnetic field decay by ambipolar
diffusion of normal or superfluid protons is given by
eq.~(\ref{eq:coreheat}),\footnote{The heating rate could be further corrected
by using $\Qheat\sim H_{\mathrm{c1}}B/4\pi\ttamb$ when the protons are
superconducting.  This would make core heating even less important at strong
fields.} where
\be
\tamb\rightarrow\ttamb = \left\{\begin{array}{ll}
\tamb & \mbox{if } T>\Tcp \\
\tambssc & \mbox{if } \Ttrsc<T<\Tcp \\
\tambirsc & \mbox{if } T<\Ttrsc
\end{array}\right. \label{eq:ttamb}
\ee
and illustrated in Fig.~\ref{fig:tdecay} for typical parameter values.
Note the removal of one power of $B$ in $\ttamb$ at $T<\Tcp$ and the strong
increase in $\ttamb$ ($\propto R_{p{\mathrm{A}}}^{-1} T^{-6}$) at low
temperatures.
Using this heating rate (that accounts for superconducting protons) in our
cooling calculations, we find that there is {\it no delay} in the onset of
core neutron superfluidity.  Onset occurs at
$\mbox{a few}\times 100\mbox{ yr}$, and the
temperature profile/evolution is very similar to the case without core
heating (see Figs.~\ref{fig:tempprofsf} and \ref{fig:tempprofcrust})
for any field $\lesssim 10^{16}\mbox{ G}$
(however, at the high field end, superconductivity is expected to be
suppressed; \citealt{baymetal69}).
This is because proton superconductivity greatly increases the timescale
for field decay, which consequently reduces $\Qheat$ and renders core
heating ineffective against cooling by strong neutrino emission.

%-------------------------------------------
%-------------------------------------------

%%%%%%%%%%%%%%%%%%%%%%%%%%%%%%%%%%%%%%%%%%%%%%%%%%%%%%%%%%%%%%%%%%%
\section{Discussion} \label{sec:discuss}

We performed detailed calculations of neutron star cooling, including the
effects of superfluidity and additional heating (due to magnetic field
decay) in the crust and core.
We find that magnetic field decay in the neutron star core cannot be the
sole source powering the high observed surface temperature of magnetars
in quiescence; the high temperatures require outer crust heating
\citep{kaminkeretal06,kaminkeretal09}.
Because of crust heating and effective thermal decoupling between the outer
crust and core, the state of matter in the core cannot be deduced from
these surface temperature measurements.
By computing the evolution of the temperature profile $T(\rho)$, we
determine the time when core neutrons first become superfluid,
i.e., when $T<\Tcnt(\rho)$.
We find that heating by field decay in the core
(with fields $\lesssim 10^{16}\mbox{ G}$) cannot balance
neutrino cooling and thus cannot maintain relatively high core temperatures
(c.f., \citealt{thompsonduncan96,dallossoetal09}).
As a result, onset of superfluidity for neutrons in the core cannot be
delayed, and neutron stars possess superfluid and superconducting
cores after a few hundred years; this does not strongly depend on the
nuclear EOS.  Since core heating is not significant,
the temperature profiles $T(\rho)$ and surface temperature evolution
$\Ts^\infty(t)$ for magnetars are just those with crust heating
(see Figs.~\ref{fig:tempprofcrust} and \ref{fig:timets}, respectively),
and crust heating does not affect the onset of core neutron superfluidity
because of thermal decoupling between the outer crust and core.

Magnetar activity may be driven by field decay in the core
\citep{thompsonduncan95,thompsonduncan96}
or by processes in the crust
\citep{thompsonduncan93,thompsonduncan95,glampedakisetal11,priceetal12}.
Although the former may still be true, it must be accompanied by heating
in the (outer) crust.
On the other hand, field evolution in the crust easily couples to surface
emission.
Thus in order to understand the high surface temperature of magnetars,
detailed studies should focus on magnetic field evolution and heating
in the crust (see, e.g., \citealt{ponsetal09,cooperkaplan10,priceetal12}).
We note that several magnetars may have very similar X-ray luminosities
\citep{durantvankerkwijk06}; this could suggest that their crustal field
strengths are similar and the field decay timescale is longer than the
age of the oldest of these sources.

Finally, the (normal versus superfluid) state of the core has important
consequences for magnetic and rotational evolution of magnetars
\citep{glampedakisandersson11},
as well as their glitching behavior and possible stellar oscillations;
studies of these effects may be effective probes of the neutron star core.
We showed that the core can be treated as being in a superfluid and
superconducting state after the neutron star is a few hundred years old.

\section*{acknowledgments}
WCGH thanks Dmitry Yakovlev for valuable comments on an early version
of the manuscript.
WCGH appreciates the use of the computer facilities at the Kavli
Institute for Particle Astrophysics and Cosmology.
WCGH and NA acknowledge support from the Science and
Technology Facilities Council (STFC) in the UK.
KG is supported by the Ram\'{o}n y Cajal Programme of the Spanish Ministerio
de Ciencia e Innovaci\'{o}n.

\bibliographystyle{mnras}

\label{lastpage}

\end{document}